     \documentclass[preprint,superscriptaddress,nofootinbib]{revtex4}
     
     \usepackage{amssymb,amsmath,graphicx,subfigure,color}
     \usepackage{rotating,multirow,dcolumn}
     \usepackage[colorlinks]{hyperref}
     \allowdisplaybreaks[4]
     \begin{document}

     \title{Searching for the toponium ${\eta}_{t}$ with the
            ${\eta}_{t}$ ${\to}$ $W^{+}W^{-}$ decay}
     \author{Yueling Yang}
     \affiliation{School of Physics,
                 Henan Normal University, Xinxiang 453007, China}
     \author{Bingbing Yang}
     \affiliation{School of Physics,
                 Henan Normal University, Xinxiang 453007, China}
     \author{Jiazhi Li}
     \affiliation{School of Physics,
                 Henan Normal University, Xinxiang 453007, China}
     \author{Zhaojie L\"{u}}
     \affiliation{School of Physics,
                 Henan Normal University, Xinxiang 453007, China}
     \author{Junfeng Sun}
     \affiliation{School of Physics,
                 Henan Normal University, Xinxiang 453007, China}

     \begin{abstract}
     Inspired by the observation of the ${\eta}_{t}$ meson
     at the LHC and the promising prospect of the ${\eta}_{t}$
     meson available at the approaching HL-LHC,
     branching ratios for the ${\eta}_{t}$ ${\to}$ $f\bar{f}$,
     $gg$, ${\gamma}{\gamma}$, $W^{+}W^{-}$, $Z^{0}Z^{0}$,
     $Z^{0}{\gamma}$ and $Z^{0}H$ decays are roughly estimated.
     It is found that tens of opposite-charge dilepton
     events from the ${\eta}_{t}$ ${\to}$ $W^{+}W^{-}$ decay and
     hundreds of events from the ${\eta}_{t}$ ${\to}$ $Z^{0}H$
     ${\to}$ ${\ell}^{+}{\ell}^{-}H$ decay using the single $Z^{0}$
     boson tagging method are expected to be accessible.
     This estimation provides a reference for future
     experimental study on the ${\eta}_{t}$ meson.

     \href{https://doi.org/10.1088/1674-1137/ae18aa}{Chin. Phys. C 50, 033101 (2026)}

     \end{abstract}
     \maketitle

     The top quark, denoted as $t$, is an extraordinary elementary
     particle in the Standard Model (SM) of particle physics.
     On one hand, the $t$ quark has the most privileged Yukawa
     coupling to the Higgs boson, and is the most massive
     elementary particle identified up to date, $m_{t}$ $=$
     $172.57{\pm}0.29$ GeV \cite{PhysRevD.110.030001}
     with a measuring accuracy better than $0.2\,\%$.
     The top quark mass is almost exactly equal to one unit of
     $v/\sqrt{2}$, where $v$ $=$ $(\sqrt{2}\,G_{F})^{-1/2}$
     $=$ $246.22$ GeV \cite{PhysRevD.110.030001}
     represents the vacuum expectation value of the
     scalar Higgs field.
     On the other hand, the considerable mass of the top quark
     facilitates an enormous phase space, which in turn results
     in a broad decay width proportional to the cube of
     the top quark mass\footnotemark[1],
     \footnotetext[1]{Neglecting the bottom quark mass and terms
     of ${\alpha}_{s}$, the partial width predicted in the SM at
     the leading order approximation is \cite{PLB.181.157},
     \begin{equation}
    {\Gamma}(t{\to}Wb) \, = \,
     \frac{ G_{F}\,m_{t}^{3} }{ 8\,{\pi}\,\sqrt{2} }\,
    {\vert}V_{tb}{\vert}^{2}\,
     \big( 1-\frac{ m_{W}^{2} }{ m_{t}^{2} } \big)^{2}\,
     \big( 1+2\,\frac{ m_{W}^{2} }{ m_{t}^{2} } \big)
     \label{eq:t2wb-decay-width},
     \end{equation}
     with the weak interaction coupling Fermi constant $G_{F}$
     ${\approx}$ $1.166\,{\times}\,10^{-5}\,{\rm GeV}^{-2}$,
     the Cabibbo-Kobayashi-Maskawa (CKM) matrix element
     $V_{tb}$ ${\approx}$ $1$, and the $W$ boson mass
     $m_{W}$ ${\approx}$ $80.4$ GeV.
     It is observed from Eq.(\ref{eq:t2wb-decay-width}) that the
     heavier the top quark mass is, the more rapidly the
     top quark width increases.
     }
     ${\Gamma}_{t}$ ${\approx}$ $1.42^{+0.19}_{-0.15}$ GeV
     \cite{PhysRevD.110.030001} with a measuring accuracy
     less than $10\,\%$, and correspondingly an instantaneous
     lifetime, ${\tau}_{t}$ $=$ $1/{\Gamma}_{t}$ ${\approx}$
     $0.5\,{\times}\,10^{-24}$~s, that is usually assumed to be
     shorter than the hadronization time.
     It was generally believed that a $t\bar{t}$ bound state
     could never be formed and observed.
     However, this conventional and prevalent view is greatly
     challenged by the recent intriguing measurements at the Large
     Hadron Collider (LHC).
     A significant excess of events close to the kinematic $t\bar{t}$
     threshold is observed independently by two LHC experiments,
     with a statistical significance of over $5$ ${\sigma}$
     discovered by the CMS experiment \cite{2507.05119,RPP.88.087801}
     and $7.7$ ${\sigma}$
     confirmed by the ATLAS experiment \cite{ALTASetat}.
     This new fascinating and mysterious resonance corresponding to
     the observed enhancement
     is preferably and consistently explained by both
     experiments with a composite color-singlet $CP$-odd
     pseudoscalar toponium, referred tp as ${\eta}_{t}(1S)$ and
     abbreviated as ${\eta}_{t}$,
     the ground $S$-wave spin-singlet ($1^{1}S_{0}$) bound
     state consisting of a top and an antitop quark $t\bar{t}$.

     Of particular interest are the properties of the
     toponium ${\eta}_{t}$.
     There are some similarities and differences between the
     toponium ${\eta}_{t}$, charmonium ${\eta}_{c}$ consisting
     of the $c\bar{c}$ quark, and bottomonium ${\eta}_{b}$ consisting
     of the $b\bar{b}$ quark.
     Based on the traditional quark model, the similarities are
     as follows:
     (a)
     they are all  the ground $S$-wave spin-singlet pseudoscalar
     heavy quarkonium with the spectroscopic notation of $1^{1}S_{0}$,
     and share the same spin, parity
     and charge-parity quantum numbers $J^{PC}$ $=$ $0^{-+}$.
     (b)
     They are all the unflavored mesons.
     Their additive intrinsic quantum numbers,
     including the baryon number, the electric charge,
     isospin, strangeness, charm, bottomness, and topness,
     are all zero.
     (c)
     Their masses are approximately the sum of their
     constituent quark masses, but just below the open-flavor threshold.
     Thus, the explicit-flavored hadronic decays,
     e.g. ${\eta}_{c}$ ${\to}$ $D\overline{D}$,
     ${\eta}_{b}$ ${\to}$ $B\overline{B}$,
     and ${\eta}_{t}$ ${\to}$ $T\overline{T}$,
     are absolutely prohibited by the law of conservation
     of energy, where the symbols of $D$, $B$ and $T$ denote
     the ground charmed, bottomed, and topped pseudoscalar
     mesons, respectively.

     The differences in the properties of the ${\eta}_{t}$,
     ${\eta}_{c}$ and ${\eta}_{b}$ include:
     (a)
     the ${\eta}_{t}$ meson consists of the heaviest top quark
     and anti-top quark, and its mass is about $m_{{\eta}_{t}}$
     ${\approx}$ $2\,m_{t}$ ${\approx}$ $343$ GeV
     \cite{2507.05119,RPP.88.087801,ALTASetat,PhysRevD.110.054032,
     PhysRevD.104.034023,PhysRevD.111.096016},
     which is two orders of magnitude greater than $m_{{\eta}_{c}}$
     $=$ $2.9839(4)$ GeV \cite{PhysRevD.110.030001} and exceeds
     35 times that of $m_{{\eta}_{b}}$ $=$ $9.3987(20)$
     GeV \cite{PhysRevD.110.030001}.
     (b)
     The toponium ${\eta}_{t}$ is supercompact\footnotemark[2],
     \footnotetext[2]{For comparison, common hadrons have
     a Bohr radius of order $r_{h}$ ${\sim}$ $1/{\Lambda}_{\rm QCD}$
     ${\sim}$ $1/200\,{\rm MeV}$ ${\sim}$ $1$ fm.
     The revolution time of toponium, estimated
     as $t$ ${\sim}$ $r/c$ \cite{PLB.181.157}, is of the same
     order of magnitude as the toponium lifetime.}
     and its Bohr radius size is of order
     $r$ ${\sim}$ $1/m_{t}\,{\alpha}_{s}(m_{t})$ ${\sim}$ $0.01$ fm,
     compared with the bottomonium size
     $r$ ${\sim}$ $1/m_{b}\,{\alpha}_{s}(m_{b})$ ${\sim}$ $0.19$ fm
     and the charmonium size
     $r$ ${\sim}$ $1/m_{c}\,{\alpha}_{s}(m_{c})$ ${\sim}$ $0.35$ fm.
     The small Bohr radius of the toponium allows probing the
     deep region of the QCD potential near the threshold where the
     strong coupling constant ${\alpha}_{s}$ is small.
     (c)
     Both components of the ${\eta}_{t}$ meson, the $t$ and $\bar{t}$
     quarks, can decay individually.
     The toponium ${\eta}_{t}$ decay width may be significantly large,
     of order ${\Gamma}_{{\eta}_{t}}$ ${\approx}$ $2\,{\Gamma}_{t}$
     ${\approx}$ $3$ GeV \cite{RPP.88.087801,ALTASetat,PhysRevD.110.054032},
     which is two orders of magnitude larger than ${\Gamma}_{{\eta}_{b,c}}$.
     The extremely large width makes the distinct ${\eta}_{t}$ meson smear
     together into a broad threshold enhancement and very hard
     to identify in experiments, which is
     in striking contrast to the ${\eta}_{b,c}$ mesons.
     In addition, the mass difference between successive toponium
     states is smaller than the toponium width, so two successive
     toponium states overlap and become indistinguishable
     \cite{APPB.15.505,PhysRevD.43.264,PhysRevD.43.1488,
     APPB.25.1837,APPB.28.2461,2412.18527,
     PhysRevD.111.096016,2506.14552}.
     (d)
     The decay configurations differ greatly between the ${\eta}_{t}$
     and ${\eta}_{b,c}$ mesons.
     Based on the calculations of Refs. \cite{PLB.60.183,
     PhysRept.41.1,PhysRevD.31.1051,PhysRevD.35.3366,PhysRevD.37.3210,
     PhysRevD.39.2668,PhysRevD.48.179,
     PhysRevD.50.3173,NuoCimA.107.2789,APPB.35.2103},
     the partial di-gluonic and di-photonic decay widths
     are inversely proportional to the square of the quarkonium
     mass\footnotemark[3],
     \footnotetext[3]{See Eq.(\ref{eq:eta2gg}) and Eq.(\ref{eq:eta2rr})}
     ${\Gamma}({\eta}_{i} {\to} gg)$ ${\propto}$
     ${\Gamma}({\eta}_{i} {\to} {\gamma}{\gamma})$ ${\propto}$
     $1/m^{2}_{{\eta}_{i}}$ with $i$ $=$ $c$, $b$ and $t$.
     The ${\eta}_{b,c}$ mesons decay predominantly through the
     chromatic and electromagnetic interactions \cite{PhysRevD.110.030001},
     while the shares of the weak interactions are negligible \cite{AHEP.2016,
     PhysLettB.751.171,PhysRevD.94.034029,IJTP.60}.
     However, the chromatic and electromagnetic decay width of
     the ${\eta}_{t}$ meson is terribly suppressed due to the huge
     mass $m_{{\eta}_{t}}$.
     The relationship of ${\Gamma}_{{\eta}_{t}}$ ${\approx}$
     $2\,{\Gamma}_{t}$ is almost equivalent to a formal announcement
     that the ${\eta}_{t}$ meson will decay overwhelmingly through
     the weak interactions.
     Therefore, the toponium decay process can be calculated reliably.

     The  observation of toponium will initiate another
     new way to study the strong interaction,
     because the heaviest top quark mass makes nonrelativistic
     approximations more reliable and perturbative QCD (pQCD)
     predictions more trustworthy.
     In fact, the quest for toponium at colliders has been
     discussed based on pQCD and potential models in many
     papers, such as Refs. \cite{NPB.183.417,
     JETP.Lett.46.525,PhysRept.167.321,ZPC.48.613,PhysRevD.43.1500,
     PhysRevD.47.56,PhysRevD.50.4341,PLB.454.137,PhysLettB.666.71,
     EPJC.60.375,JHEP.2010.09.034,PLB.866.139510,PLB.866.139532,
     EPJC.85.157,PhysRevD.111.114020}.
     The spin-triplet $n^{3}S_{1}$ vector toponium states,
     ${\Theta}(nS)$ mesons, can be directly produced at the
     hadron colliders through the $q\bar{q}$ annihilation processes
     and future $e^{+}e^{-}$ colliders
     \cite{ZPC.48.613,PhysRevD.43.1500,PhysRevD.47.56,
     PhysRevD.50.4341,PLB.454.137}, such as the CEPC \cite{CEPC}
     and FCC-ee \cite{FCCee}.
     The spin-singlet $n^{1}S_{0}$ pseudoscalar toponium states,
     ${\eta}_{t}(nS)$ mesons, are promisingly accessible at hadron colliders
     through the gluon-gluon color-singlet fusion processes
     \cite{PhysLettB.666.71,EPJC.60.375,JHEP.2010.09.034,
     PLB.866.139510,PLB.866.139532}.
     The pQCD theoretical estimation on the ${\eta}_{t}$ production
     cross section with the gluon fusion mechanism,
     including state-of-the-art higher order QCD corrections, is
     ${\sigma}({\eta}_{t})$ $=$ 
     $6.43$ (or $7.54$) pb \cite{PhysRevD.104.034023}
     at the centre-of-mass energy $\sqrt{s}$ $=$ $13$ (or 14) TeV
     at the LHC, which is marginally in agreement with the measured
     cross section at $\sqrt{s}$ $=$ $13$ TeV with an integrated
     luminosity of ${\sim}$ $140$ ${\rm fb}^{-1}$, {\it e.g.},
     $8.8^{+1.2}_{-1.4}$ pb with the CMS detector
     \cite{RPP.88.087801} and $9.0{\pm}1.3$ pb with
     the ATLAS detector \cite{ALTASetat}.
     With an integrated luminosity of about 3 ${\rm ab}^{-1}$
     at $14$ TeV over 10 years of operation of the HL-LHC
     \cite{EPJST.228.1109}, more than $2{\times}10^{7}$
     ${\eta}_{t}$ mesons are expected to be available in the
     future, offering valuable opportunities and
     promising prospects to discover and study the ${\eta}_{t}$
     meson at high-energy and high-luminosity experiments.

     The low near-threshold production ratio relative to the non-resonant
     $t\bar{t}$ production cross section\footnotemark[4]
     \footnotetext[4]{It is theoretically estimated that the
     ${\eta}_{t}$ meson production contributes to $0.79\,\%$ of
     the total non-resonant $t\bar{t}$ production cross section
     at $13$ TeV at the LHC \cite{PhysRevD.104.034023}.
     The ratio of the production cross section
     ${\sigma}({\eta}_{t})/{\sigma}(t\bar{t})$
     is measured to be
     $8.8^{+1.2}_{-1.4}\,{\rm pb}/833.9^{+20.5}_{-30.0}\,{\rm pb}$
     ${\sim}$ $1.06(17)\,\%$
     by the CMS group \cite{RPP.88.087801}
     and $9.0{\pm}1.3\,{\rm pb}/833.9^{+37.4}_{-43.0}\,{\rm pb}$
     ${\sim}$ $1.08(17)\,\%$
     by the ATLAS group \cite{ALTASetat}.}
     and the large decay width make the
     identification of the paratoponium ${\eta}_{t}$ extremely
     difficult against the complicated $t\bar{t}$ muddy
     entanglement background.
     It is indisputable that the dominant ${\eta}_{t}$ decay mode
     is the intrinsic decays of the constituent top and anti-top
     quarks.
     With the obvious hierarchy relations among the CKM matrix
     elements, ${\vert} V_{tb} {\vert}$ ${\gg}$
     ${\vert} V_{ts} {\vert}$ ${\gg}$ ${\vert} V_{td} {\vert}$,
     the top quark decay, almost exclusively into a real $W$ boson
     and a bottom quark, $t$ ${\to}$ $W^{+}b$, is the dominant
     channel, where the $b$-jets can be distinguished experimentally
     from other jets due to the long lifetime of the bottom quarks
     and relativistic time dilation.
     The decay rate is ${\Gamma}(t{\to}Wb)/{\Gamma}(t{\to}Wq)$
     $=$ $(95.7{\pm}3.4)\,\%$ \cite{PhysRevD.110.030001}, where $q$
     denotes all the weak-isospin down-type quarks $q$ $=$ $d$,
     $s$, and $b$.
     The ${\eta}_{t}$ meson decay is predominantly induced by the
     $t$ ${\to}$ $W^{+}b$ decay, {\it i.e.},
     ${\eta}_{t}$ ${\to}$ $\overline{T}$ $+$ $W^{+}$ $+$ $b$
     (or $T$ $+$ $W^{-}$ $+$ $\bar{b}$) \cite{NPB.198.71}
     if the topped $T$ hadrons could instantaneously exist
     \cite{2508.03422,2508.17646},
     followed immediately by a complex cascade decay series
     of the $T$ hadrons and $W$ bosons.
     In principle, the ${\eta}_{t}$ meson and the non-resonant
     $t\bar{t}$ pair will have the same final states
     $W^{+}bW^{-}\bar{b}$.
     Experimentally, based on whether the final states of the
     $W^{\pm}$ boson decays are the leptons or quarks,
     the ${\eta}_{t}$ and non-resonant $t\bar{t}$ event
     reconstruction can be divided
     into three classes  \cite{PhysRevD.110.030001}:
     (a)
     the dilepton channels, where both $W$ bosons decay into leptons\footnotemark[5],
     with a ratio of $10.5\%$,
     \footnotetext[5]{The dilepton channels where both $W$
     bosons decay into $e\,{\nu}_{e}$ or ${\mu}\,{\nu}_{\mu}$
     have a share of ${\sim}$ $5\%$, but with relatively little
     background. The ${\tau}$ channels where
     one or both $W$ bosons decay into ${\tau}\,{\nu}_{\tau}$
     are very difficult to identify with present detectors
     due to the additional neutrinos from the ${\tau}$ decays.}
     (b)
     the hadronic channels, where both $W$ bosons decay into quarks,
     with a ratio of $45.7\%$,
     and (c)
     the lepton+jets channels, where one $W$ boson decay into quarks,
     and the other $W$ boson decays into leptons,
     with a ratio of $43.8\%$.
     Anyhow, the decay products of the ${\eta}_{t}$ meson and the
     $t\bar{t}$ pair are miscellaneous, resulting in
     the horrible complexity of kinematics and dynamics.

     The toponium decays have a variety of interesting final
     state topologies compared with the ${\eta}_{b,c}$
     meson decays.
     Besides the dominant decay ${\eta}_{t}$ ${\to}$ $W^{+}bW^{-}\bar{b}$,
     there is also conventional signals from the ${\eta}_{t}$ ${\to}$
     $gg$, ${\gamma}{\gamma}$, $f\bar{f}$ decays.
     Moreover, the potentially interesting signals are the ${\eta}_{t}$ meson
     decay into the electroweak gauge boson pairs $W^{+}W^{-}$,
     $Z^{0}Z^{0}$, $Z^{0}{\gamma}$ and also into $Z^{0}H$,
     where the $W^{\pm}$ and $Z^{0}$ gauge bosons devour
     three degrees of the freedom of the Higgs field and acquire masses.
     These final states containing the on-shell $W^{\pm}$ or $Z^{0}$
     or Higgs particles are kinematically inaccessible for the
     ${\eta}_{b,c}$ meson decays due to the energy conservation.
     The ${\eta}_{t}$ ${\to}$ ${\gamma}H$ decay is not allowed
     by the $C$-parity conservation law.
     Due to the Majorana character of the color-singlet Higgs scalar
     particle with $J^{PC}$ $=$ $0^{++}$, the ${\eta}_{t}$ meson decay
     into two identical Higgs particles, ${\eta}_{t}$ ${\to}$ $HH$,
     is forbidden by the Bose-Einstein statistics and the $CP$
     conservation law \cite{PhysRevD.37.3210}.
     The lowest-order expressions of these partial widths are listed
     as follows
     \cite{PhysRept.167.321,PLB.60.183,PhysRept.41.1,PhysRevD.31.1051,
     PhysRevD.35.3366,PhysRevD.37.3210,PhysRevD.39.2668,PhysRevD.48.179,
     PhysRevD.50.3173,NuoCimA.107.2789,APPB.35.2103,2506.14552}. %
     \begin{equation}
    {\Gamma}({\eta}_{t}{\to}gg) \, = \,
     \frac{8}{3} \, {\alpha}_{s}^{2} \,
     \frac{ {\vert} R_{S}(0) {\vert}^{2} }
          { m^{2}_{{\eta}_{t}} }
     \label{eq:eta2gg},
     \end{equation}
     \begin{equation}
    {\Gamma}({\eta}_{t}{\to}{\gamma}{\gamma}) \, = \,
     \frac{64}{27} \, {\alpha}_{\rm em}^{2} \,
     \frac{ {\vert} R_{S}(0) {\vert}^{2} }
          { m^{2}_{{\eta}_{t}} }
     \label{eq:eta2rr},
     \end{equation}
     \begin{equation}
    {\Gamma}({\eta}_{t}{\to}f\bar{f}) \, = \,
     N_{f}\, \frac{3\, {\alpha}_{Z}^{2} }{32} \,
     \frac{ x_{f} }
          { x^{2}_{Z} }\, {\lambda}^{1/2}(1,x_{f},x_{f})
     \frac{ {\vert} R_{S}(0) {\vert}^{2} }
          { m^{2}_{{\eta}_{t}} }
     \label{eq:eta2ff},
     \end{equation}
     \begin{equation}
    {\Gamma}({\eta}_{t}{\to}W^{+}W^{-}) \, = \,
     \frac{3\, {\alpha}_{W}^{2} }{ 2  } \,
    {\lambda}^{-1/2}(1,x_{W},x_{W}) \,
     \frac{ {\vert} R_{S}(0) {\vert}^{2} }
          { m^{2}_{{\eta}_{t}} }
     \label{eq:eta2ww},
     \end{equation}
     \begin{equation}
    {\Gamma}({\eta}_{t}{\to}Z^{0}Z^{0}) \, = \,
     \frac{ {\alpha}_{Z}^{2} }{ 432  }\,
     \frac{ (9-24\,{\sin}^{2}{\theta}_{W}+32\,{\sin}^{4}{\theta}_{W})^2 }
          { (1-2\,x_{Z})^{2} \, {\lambda}^{-3/2}(1,x_{Z},x_{Z}) }\,
     \frac{ {\vert} R_{S}(0) {\vert}^{2} }
          { m^{2}_{{\eta}_{t}} }
     \label{eq:eta2zz},
     \end{equation}
     \begin{equation}
    {\Gamma}({\eta}_{t}{\to}Z^{0}{\gamma}) \, = \,
     \frac{ 2\, {\alpha}_{\rm em}\,{\alpha}_{Z} }
          { 27  }\,
     \frac{ (3-8\,{\sin}^{2}{\theta}_{W})^2 }
          { {\lambda}^{-1/2}(1,x_{Z},0)  }\,
     \frac{ {\vert} R_{S}(0) {\vert}^{2} }
          { m^{2}_{{\eta}_{t}} }
     \label{eq:eta2zr},
     \end{equation}
     \begin{equation}
    {\Gamma}({\eta}_{t}{\to}Z^{0}H) \, = \,
     \frac{ 3\, {\alpha}_{Z}^{2} }
          { 64  }\,
     \frac{ {\lambda}^{3/2}(1,x_{Z},x_{H}) }
          { x_{Z}^{2}  }\,
     \frac{ {\vert} R_{S}(0) {\vert}^{2} }
          { m^{2}_{{\eta}_{t}} }
     \label{eq:eta2zh},
     \end{equation}
     where the strong coupling constant ${\alpha}_{s}$,
     the electromagnetic fine-structure constant ${\alpha}_{\rm em}$,
     the electroweak coupling factors ${\alpha}_{Z}$ $=$
     ${\alpha}_{\rm em}/({\sin}^{2}{\theta}_{W}\,{\cos}^{2}{\theta}_{W})$
     and ${\alpha}_{W}$ $=$ ${\alpha}_{\rm em}/{\sin}^{2}{\theta}_{W}$,
     the ratio of the mass square $x_{i}$ $=$ $m_{i}^{2}/m^{2}_{{\eta}_{t}}$,
     and the color number $N_{f}$ in Eq.(\ref{eq:eta2ff}) is equal to
     $1$ for the leptons and $3$ for the quarks.
     ${\lambda}(a,b,c)$ $=$ $a^{2}$ $+$ $b^{2}$ $+$ $c^{2}$ $-$
     $2\,a\,b$ $-$ $2\,b\,c$ $-$ $2\,c\,a$ is the K\"{a}ll\'{e}n function.
     $R_{S}(0)$ is the radial wave function of the $S$ wave state
     evaluated at the origin $r$ $=$ $0$, and can approximately
     be obtained with the solution of the Schr\"{o}dinger equation
     with a phenomenological potential.
     The above partial decay widths are proportional to
     ${\vert} R_{S}(0) {\vert}^{2} / m^{2}_{{\eta}_{t}}$.
     The effects arising from the higher-order QCD corrections
     seem to be limited due to the
     small coupling ${\alpha}_{s}(m_{t})$.
     The total decay width ${\Gamma}_{{\eta}_{t}}$ 
     should be dominated by the single-top-quark electroweak  decay,
     which is independent of the value of the wave function
     at the origin.

     The toponium strongly resembles a non-relativistic positronium.
     The interactions between the $t$ and $\bar{t}$ quarks at the
     short distances should be governed by electroweak and
     perturbative QCD effects \cite{ZPC.48.613}.
     For the superheavy quarkonia with a sub-femtometer Bohr radius,
     a non-relativistic treatment of the interquark potential
     becomes possible due to the asymptotic freedom of QCD.
     The potential is usually written as a short-distance
     part $V_{S}$ plus a long-distance part $V_{L}$,
     where the $V_{S}$ part arises from the gluon-exchange
     interaction between quarks, and
     the $V_{L}$ part is motivated by confinement \cite{NPB.344.1}.
     It is recognized phenomenologically that the Bohr radius
     of the toponium is sufficiently deep in $V_{S}$
     and thus sufficiently far away from the confinement part
     of the potential.
     The radial wave functions at the origin should be predominantly
     determined by the short-distance potential $V_{S}$ with the
     spherically symmetric and static Coulomb-like form
     \cite{ZPC.48.613,PhysRevD.43.1500,PhysRevD.47.56,
     PhysRevD.50.4341,NPB.344.1}.
     \begin{equation}
     V(r) \, = \, - \, C_{F}\, \frac{{\alpha}_{s}}{r}
     \label{eq:Coulomb-potentials},
     \end{equation}
     where $C_{F}$ $=$ $4/3$.
     The general expression for the radial wave functions at the origin
     of the $S$ states is given by \cite{NPB.344.1},
     \begin{equation}
    {\vert} R_{nS}(0) {\vert}^{2} \, = \,
     \frac{4}{n^{3}}\, \big( C_{F}\, {\alpha}_{s}\, {\mu}_{Q} \big)^{3}
     \label{eq:wf-ns-r0},
     \end{equation}
     where the reduced mass ${\mu}_{Q}$ $=$ $m_{t}/2$ for the
     ${\eta}_{t}(nS)$ mesons.

     \begin{table}[ht]
     \caption{The mass of particles and physical constants are taken from
              Ref. \cite{PhysRevD.110.030001},
              where their central values are regarded as the default
              inputs unless otherwise specified.}
     \label{tab:input}
     \begin{ruledtabular}
     \begin{tabular}{llll}
       Mass of leptons
     & $ m_{\tau} $ $=$ $ 1776.93 (9) $ MeV,
     & $ m_{\mu} $ $=$ $ 105.658 $ MeV,
     \\
        Mass of quarks
     & $ m_{t} $ $=$ $ 172.57 (29) $ GeV,
     & $ m_{b} $ $=$ $ 4.78 (6) $ GeV,
     & $ m_{c} $ $=$ $ 1.67 (7) $ GeV,
     \\
       Mass of bosons
     & $ m_{H} $ $=$ $ 125.20 (11) $ GeV,
     & $ m_{Z} $ $=$ $ 91.1880 (20) $ GeV,
     & $ m_{W} $ $=$ $ 80.3692 (133) $ GeV,
     \\
       Physical constant
     & $ {\alpha}_{\rm em}(m_{W}) $ $=$ $ 1/128 $,
     & $ {\alpha}_{s}(m_{Z}) $ $=$ $ 0.1180(9) $,
     & $ {\sin}^{2}{\theta}_{W} $ $=$ $ 0.23129(4) $.
     \end{tabular}
     \end{ruledtabular}
     \end{table}
     \begin{table}[h]
     \caption{The possible partial widths (${\Gamma}_{i}$), and
        branching ratios (${\cal B}r_{i}$ $=$ ${\Gamma}_{i}/{\Gamma}_{{\eta}_{t}}$),
        and the event numbers ($N_{i}$ $=$ ${\cal B}r_{i}{\times}N_{{\eta}_{t}}$)
        of the ${\eta}_{t}$ decay, with the full
        width ${\Gamma}_{{\eta}_{t}}$ $=$ $3$ GeV
        and $N_{{\eta}_{t}}$ $=$ $2{\times}10^{7}$.}
     \label{tab:wd-etat-1s}
     \begin{ruledtabular}
     \begin{tabular}{crcr}
       decay mode & \multicolumn{1}{c}{${\Gamma}_{i}$}
     & ${\cal B}r_{i}$ & \multicolumn{1}{c}{$N_{i}$}  \\ \hline
     ${\mu}^{+}\,{\mu}^{-}$   & $0.22$   ~eV & $ 7.39 \, {\times}\, 10^{-11}$ & $ 0~~ $ \\
     ${\tau}^{+}\,{\tau}^{-}$ & $62.68$  ~eV & $ 2.09 \, {\times}\, 10^{-8} $ & $ 0.4 $ \\
     $c\,\bar{c}$             & $166.09$ ~eV & $ 5.54 \, {\times}\, 10^{-8} $ & $ 1~~ $ \\
     $b\,\bar{b}$             & $ 1.36$  keV & $ 4.53 \, {\times}\, 10^{-7} $ & $ 9~~ $ \\
     $g\,g$                & $ 1989.08$  keV & $ 6.63 \, {\times}\, 10^{-4} $ & $ 13261~~ $ \\
     ${\gamma}\,{\gamma}$     & $ 9.32$  keV & $ 3.11 \, {\times}\, 10^{-6} $ & $ 62~~ $ \\
     $W^{+}\,W^{-}$           & $97.45$  keV & $ 3.25 \, {\times}\, 10^{-5} $ & $ 650~~ $ \\
     $Z^{0}\,Z^{0}$           & $ 6.32$  keV & $ 3.11 \, {\times}\, 10^{-6} $ & $ 42~~ $ \\
     $Z^{0}\,{\gamma}$        & $ 2.01$  keV & $ 6.71 \, {\times}\, 10^{-7} $ & $ 13~~ $ \\
     $Z^{0}\,H$               & $537.43$ keV & $ 1.79 \, {\times}\, 10^{-4} $ & $ 3583~~ $
     \end{tabular}
     \end{ruledtabular}
     \end{table}

     With the above formula for the partial widths in
     Eq.(\ref{eq:eta2gg})---Eq.(\ref{eq:eta2zh}),
     the radial wave function in Eq.(\ref{eq:wf-ns-r0}),
     and the inputs in Table \ref{tab:input},
     the estimated partial widths and branching ratios of
     the ${\eta}_{t}(1S)$ meson decay into different final
     states are listed in Table \ref{tab:wd-etat-1s},
     which is consistent with those obtained in Ref. \cite{2506.14552}
     if using a relatively large coupling constant
     ${\alpha}_{s}$ $=$ $0.189$ and small total width
     ${\Gamma}_{{\eta}_{t}}$ $=$ $2.84$ GeV.
     Here it should be pointed out that the numbers in
     Table \ref{tab:wd-etat-1s} are only a rough estimate.
     For example, it has been shown \cite{PhysRevD.37.3210,
     PhysRevD.48.179,PhysRevD.50.3173,NuoCimA.107.2789,APPB.35.2103}
     that the QCD corrections to the partial
     width for the ${\eta}_{t}$ ${\to}$ $gg$, ${\gamma}{\gamma}$
     decays could reach up to ${\sim}$ $10\,\%$.
     In addition, the mass $m_{{\eta}_{t}}$ and the decay
     width ${\Gamma}_{{\eta}_{t}}$ have not been determined
     experimentally.
     Furthermore, the theoretical uncertainties from
     the top quark mass,
     the higher order electroweak correction effects,
     the different forms of the radial wave functions,
     the threshold effects,
     the interquark potential model dependence, and so on,
     which will have much influence on the results,
     are not considered carefully here.
     Notwithstanding, the estimated numbers in
     Table \ref{tab:wd-etat-1s} have a certain reference significance
     in investigating the ${\eta}_{t}$ decays.
     It is seen from Table \ref{tab:wd-etat-1s} that
     (1)
     among the traditional decay modes, the ${\eta}_{t}$ ${\to}$
     $gg$ decay is the dominant one.
     The chromo gluons will convert into the quark and antiquark pairs
     and finally fragment into various hadrons after a
     complicated hadronization process.
     So, the ${\eta}_{t}$ ${\to}$ $gg$ decay would be obscured
     by the strong interaction backgrounds.
     (2)
     For the ${\eta}_{t}$ ${\to}$ ${\gamma}{\gamma}$ decays,
     the high energy photons should be efficiently reconstructed from
     their energy deposits in the calorimeter at the LHC.
     The photon identification efficiency exceeds $95\%$ with the
     ATLAS experiment
     for the transverse energy $100$ GeV $<$ $E_{T}$ $<$ $200$ GeV
     in the pseudorapidity range of ${\vert}{\eta}{\vert}$ $<$ $2.37$
     \cite{JHEP.2024.05.162}.
     (3)
     The branching ratios for the ${\eta}_{t}$ ${\to}$ $f\bar{f}$
     decays, being directly proportional to the square of the
     fermion mass as in Eq.(\ref{eq:eta2ff}), are very small,
     and might be beyond the detectability limits when considering
     the complex backgrounds at hadron collisions.
     (4)
     The ${\eta}_{t}$ ${\to}$ $Z^{0}Z^{0}$, $Z^{0}{\gamma}$,
     $Z^{0}H$ channels may signal the ${\eta}_{t}$ existence.
     The $Z^{0}$ boson is usually and effectively reconstructed
     through its decays into the $e^{+}e^{-}$ or ${\mu}^{+}{\mu}^{-}$
     pairs at the ATLAS and CMS experiments,
     where the leptons provide a clean signature and ensure high
     trigger efficiency and good invariant mass resolution
     \cite{PhysRevLett.132.021803}.
     However, the event reconstruction from all the final leptons
     seems to be exceedingly difficult or inaccessible because
     the branching
     ratios for the ${\eta}_{t}$ ${\to}$ $Z^{0}Z^{0}$ ($Z^{0}{\gamma}$)
     ${\to}$ ${\ell}^{+}{\ell}^{-}{\ell}^{{\prime}+}{\ell}^{{\prime}-}$
     (${\ell}^{+}{\ell}^{-}{\gamma}$) decays and the ${\eta}_{t}$ ${\to}$
     $Z^{0}({\to}{\ell}^{+}{\ell}^{-}) \,H({\to}{\mu}^{+}{\mu}^{-})$ decay
     are about $10^{-9}$,
     using ${\cal B}r(Z^{0}{\to}{\ell}^{+}{\ell}^{-})$ ${\sim}$ $3.4\%$
     and ${\cal B}r(H{\to}{\mu}^{+}{\mu}^{-})$ ${\sim}$ $2.6{\times}10^{-4}$
     \cite{PhysRevD.110.030001}.
     The experimental research on the ${\eta}_{t}$ ${\to}$
     $Z^{0}({\to}{\ell}^{+}{\ell}^{-}) \, H({\to}{\tau}^{+}{\tau}^{-})$
     decay is strongly influenced by the additional invisible
     neutrinos from the ${\tau}$ decays.
     Perhaps the single $Z^{0}$ boson tagging method could be used
     to search for and explore the ${\eta}_{t}$ ${\to}$ $Z^{0}H$ decay,
     then more than $200$ events of the ${\eta}_{t}$ ${\to}$ $Z^{0}H$
     ${\to}$ ${\ell}^{+}{\ell}^{-}H$ decay (with ${\ell}$ $=$ $e$ and ${\mu}$)
     are expected to be observable.
     (5)
     For the ${\eta}_{t}$ ${\to}$ $W^{+}W^{-}$ decay,
     at least two $b$-jets are less than the decay products
     $W^{+}bW^{-}\bar{b}$ of the $t\bar{t}$ pair and
     the ${\eta}_{t}$ meson.
     In addition, the charged $W^{\pm}$ bosons
     are back-to-back in the rest frame of the ${\eta}_{t}$
     meson, and have definite energy and momenta, which will
     help to recognize unambiguous signals and minimize
     chaotic backgrounds.
     The identification technology of the $W$ bosons
     is very sophisticated at experiments.
     Under certain circumstances, the single $W$ boson tagging
     analysis methods can be used to improve the reconstruction
     efficiency.
     Considering the branching ratio for the leptonic decays of
     the $W$ bosons ${\cal B}r(W^{+}{\to}{\ell}^{+}{\nu}_{\ell})$
     ${\sim}$ $11\%$ \cite{PhysRevD.110.030001},
     it is expected to observe some $30$ opposite-charge dilepton events where
     both $W$ bosons decay into $e{\nu}_{e}$ or ${\mu}{\nu}_{\mu}$,
     and some $100$ lepton+jets events where one $W$ boson decays
     into $e{\nu}_{e}$ or ${\mu}{\nu}_{\mu}$ and the other $W$
     boson decays into quarks.
     Of course, if using the branching ratio
     ${\cal B}r({\eta}_{t}{\to}W^{+}W^{-})$ $=$ $2.42{\times}10^{-4}$
     estimated by Ref. \cite{2506.14552},
     the events of the ${\eta}_{t}$ ${\to}$ $W^{+}W^{-}$ decay will increase sevenfold.
     The ${\eta}_{t}$ ${\to}$ $W^{+}W^{-}$ decay provides a specific
     and feasible process to identify the ${\eta}_{t}$ meson.

     In summary, the intriguing paratoponium ${\eta}_{t}$, as one of
     the most compact bound states consisting of the $t\bar{t}$ pair
     and beyond conventional imagination, has been observed by
     both the CMS and ATLAS groups with a statistical significance of
     over $5\,{\sigma}$ now.
     The properties of the ${\eta}_{t}$ meson, including its mass,
     decay width and modes, production cross section, and
     so forth, will become a focus of attention for many
     particle physicists.
     Due to the large mass of the ${\eta}_{t}$ meson, some unusual
     and characteristic final states, such as $W^{+}W^{-}$, $Z^{0}Z^{0}$
     and $Z^{0}H$, can be accessible with the ${\eta}_{t}$ decays.
     Encouraged by a promising prospect of more than $2{\times}10^{7}$
     ${\eta}_{t}$ mesons available at the forthcoming HL-LHC,
     a rough magnitude order estimation on the branching ratios for
     the two-body ${\eta}_{t}$ ${\to}$ $f\bar{f}$, $gg$,
     ${\gamma}{\gamma}$, $W^{+}W^{-}$, $Z^{0}Z^{0}$,
     $Z^{0}{\gamma}$ and $Z^{0}H$ decays is calculated.
     If considering the $Z^{0}$ and $W^{\pm}$ boson reconstruction
     via the pure lepton flavors, tens of opposite-charge dilepton
     events from the ${\eta}_{t}$ ${\to}$ $W^{+}W^{-}$ decay and
     hundreds of events from the ${\eta}_{t}$ ${\to}$ $Z^{0}H$
     ${\to}$ ${\ell}^{+}{\ell}^{-}H$ decay using the single $Z^{0}$
     boson tagging method, are expected to be exploited.
     We wish that our estimation of the ${\eta}_{t}$ decays can
     provide a ready reference for the future experimental
     probe and study of the ${\eta}_{t}$ meson.

     \section*{Acknowledgments}
     The work is supported by the National Natural Science Foundation
     of China (Grant Nos. 12275068, 12275067),
     the National Key R\&D Program of China (Grant No. 2023YFA1606000),
     Natural Science Foundation of Henan Province
     (Grant Nos. 252300421491, 242300420250),
     and the Science and Technology R\&D Program Joint
     Fund Project of Henan Province (Grant No. 225200810030) and
     Science and Technology Innovation Leading Talent Support
     Program of Henan Province (Grant No. 254200510039).

     

     \end{document}